\documentclass{aa}
\usepackage{graphicx}
\usepackage{natbib}
\begin{document}

   \title{A standard stellar library for evolutionary synthesis.}
   \subtitle{III. Metallicity calibration}

   \author{P. Westera
          \inst{1}
          \and
          T. Lejeune\inst{2}
          \and
          R. Buser\inst{1}
          \and
          F. Cuisinier\inst{3}
          \and
          G. Bruzual A.\inst{4}
          }

   \offprints{R. Buser}

   \institute{Astronomisches Institut der Universit\"at Basel,
             Venusstrasse 7, CH-4102 Binningen, Switzerland\\
             \email{westera@astro.unibas.ch, buser@astro.unibas.ch}
         \and
             Observat\'orio Astron\'omico da Universidade de Coimbra, Portugal\\
             \email{lejeune@mat.uc.pt}
         \and
             Depto. de Astronomia, Universidade Federal do Rio de Janeiro, Brazil\\
             \email{cuisinie@sun1.ov.ufrj.br}
         \and
             Centro de Investigaciones de Astronom\'{\i}a, M\'erida, Venezuela\\
             \email{bruzual@cida.ve}
             }

   \date{Received <date>; accepted <date>}

   \abstract{
   We extend the colour calibration of the widely used BaSeL standard
   stellar library (Lejeune, Cuisinier, \& Buser 1997, 1998)
   to non-solar metallicities, down to ${\rm [Fe/H]} \sim -2.0$ dex.
   Surprisingly, we find that at the present epoch it is virtually
   {\em impossible to establish a unique calibration of $UBVRIJHKL$
   colours in terms of stellar metallicity ${\rm [Fe/H]}$ which is consistent
   simultaneously with both colour-temperature relations and colour-absolute
   magnitude diagrams (CMDs) based on observed globular cluster photometry
   data and on published, currently popular standard stellar evolutionary
   tracks and isochrones.} \\
   The problem appears to be related to the long-standing incompleteness
   in our understanding of convection in late-type stellar evolution, but
   is also due to a serious lack of relevant observational calibration
   data that would help resolve, or at least further significant progress
   towards resolving this issue. \\
   In view of the most important applications of the BaSeL library, we here
   propose two different metallicity calibration versions: (1) the ''WLBC 99''
   library, which consistently matches empirical colour-temperature relations
   and which, therefore, should make an ideal tool for the study of individual
   stars; and (2), the ''PADOVA 2000 ''library, which provides isochrones from the
   Padova 2000 grid \citep{padova} that successfully reproduce Galactic
   globular-cluster colour-absolute magnitude diagrams and which thus should
   prove particularly useful for studies of collective phenomena in stellar
   populations in clusters and galaxies.
   }

   \maketitle
%

\section{Introduction}

   As present grids of theoretical spectral energy distributions (SEDs)
   suffer from intrinsic
   inhomogeneities and incompleteness and show large systematic discrepancies
   with empirical calibrations due to unavailable molecular opacity
   \citep[see][hereafter Paper I]{paperi}, we have
   undertaken the construction of a comprehensive combined library of realistic
   stellar flux distributions. Empirical ${T_{\rm eff}}$-colour relations in
   $UBVRIJHKL$ photometry are used to adjust the spectra using an algorithm
   developed by Cuisinier et al. \citep[see][Paper I]{buser}. \\
   The current state of the art is the following: The semi-empirical BaSeL
   (Basel Stellar Library) 2.2 SED library \citep[][hereafter Paper II]{paperii}
   has been widely used successfully in different areas \citep{ortolani,
   kauffmann,westera,gonzalez,origlia,leitherer,bruz,brocato,maraston,barmby,
   liu,oblak,lotz,molla,nikolaev,kong,yoon,marleau,vacca,lastennet,kotilainen,
   fricke}.
   However, as a result of being calibrated from solar metallicity data only,
   BaSeL 2.2 still has its weaknesses at low metallicities
   (${\rm [Fe/H]} < -1$), especially in the ultraviolet ($U-B$) and the
   infrared ($V-K$, $J-H$, $H-K$, $J-K$, $K-L$).
   In these colours and at these lower metallicities, synthetical globular
   cluster CMDs appear too blue using the BaSeL 2.2 semi-empirical library.
   Furthermore, the transition between dwarfs and giants produces a
   discontinuity in some colours (for more details, see Papers I and II as well
   as T. Lejeune's thesis (1997)). \\
   The purpose of the present paper is to remove the weaknesses of the
   BaSeL 2.2 semi-empirical library by extending the colour-calibration to low
   metallicities, and to create a library that reproduces empirical
   colour-temperature relations and globular cluster CMDs (using existing grids
   of isochrones) at all metallicities. \\
   The outline of the paper is the following: In section 2, we discuss the
   compilation and properties of the calibration data, i. e. globular cluster
   CMDs and empirical colour-${T_{\rm eff}}$ relations.
   In section 3, we briefly describe the calibration algorithm and the changes
   made relative to the previous algorithm (described in Paper I). As a result,
   we present the BaSeL 3.1 ''WLBC 99'' SED library, which is able to reproduce
   empirical colour-temperature relations in all ($UBVRIJHKL$) colours, and is
   therefore projected to be a powerful tool for studies of individual stars.
   Unfortunately, the library doesn't provide any improvements in the representation
   of globular cluster CMDs, and it even proved impossible to provide a library that
   satisfies both requirements at the same time.
   As a pragmatic solution, we also produce an application-oriented library,
   the BaSeL ''Padova 2000'' library, which, if used along with the Padova 2000
   isochrones, successfully reproduces globular cluster CMDs at all levels of
   metallicity; this version of the library is presented and discussed in
   section 4. The conclusions are summarised in section 5, where an outlook
   on future work and on a first application can also be found. \\
   A more detailed description of this work is given in \citet{diss}.


\section{Compilation and discussion of calibration data}
\label{compilation}
   \begin{figure*}
\includegraphics[width=\textwidth]{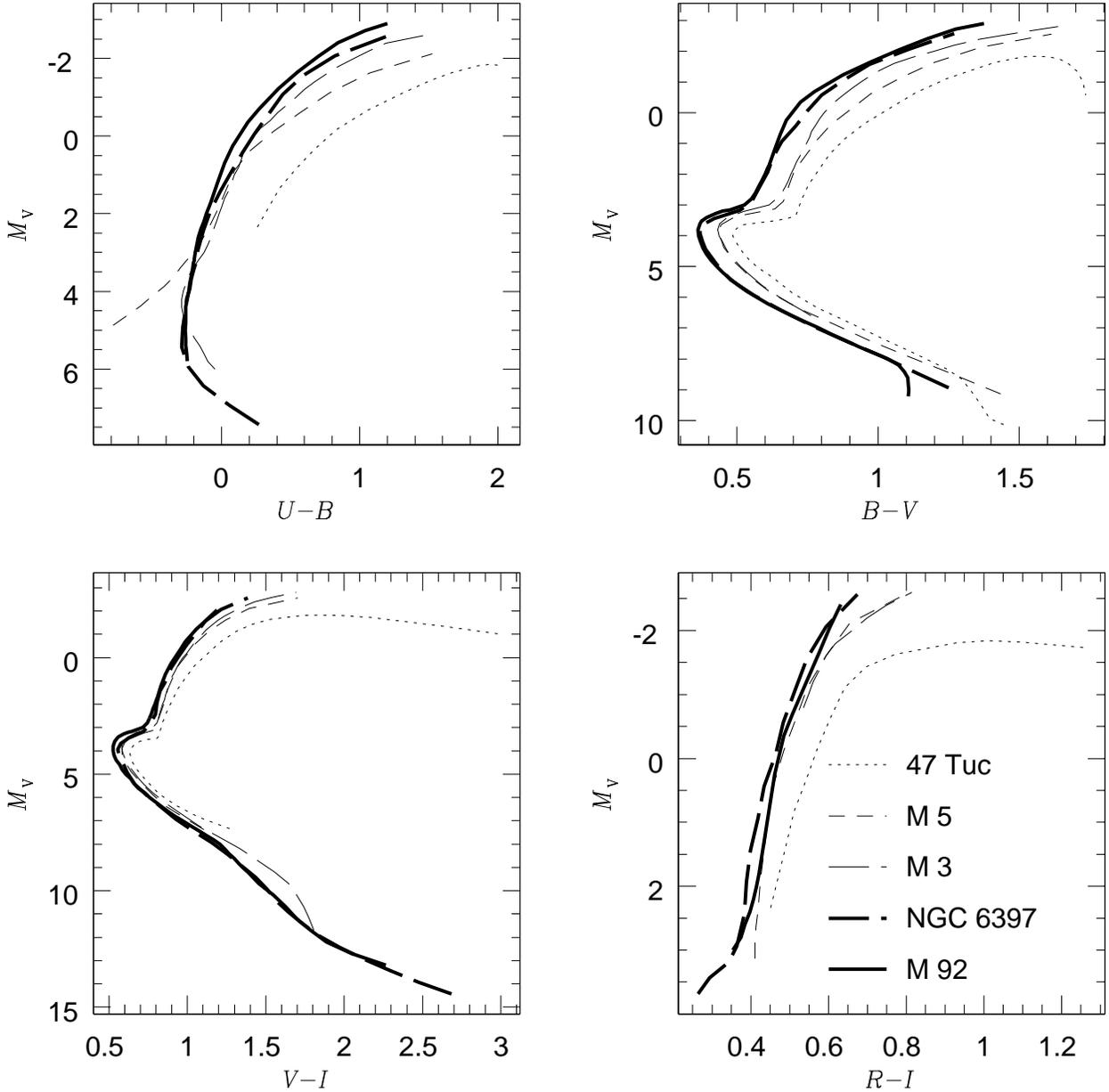}
   \caption{Empirical ($UBVRI$) fiducial lines in the CMD of the globular
               clusters 47 Tuc, M5, M3, NGC6397 and M92.}
    \label{CMDsa}
    \end{figure*}
   \begin{figure*}
\includegraphics[width=\textwidth]{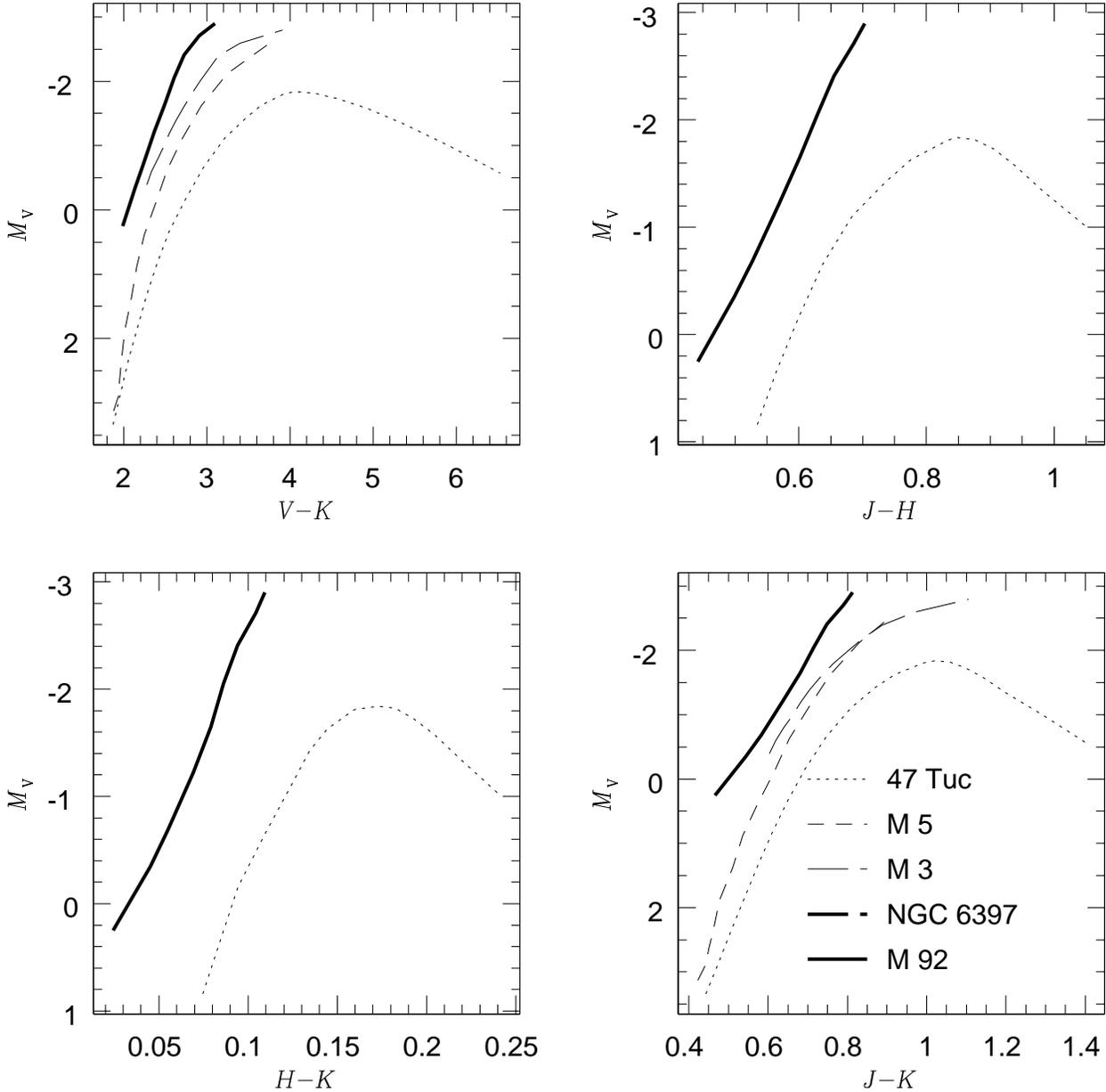}
   \caption{Empirical infrared fiducial lines in the CMD of the globular
               clusters 47 Tuc, M5, M3, NGC6397 and M92. For $V-K$ and $J-K$,
               no NGC6397 data were available, and for $J-H$ and $H-K$, only
               47 Tuc and M92 data were found.}
    \label{CMDsb}
    \end{figure*}
%
   \begin{table*}
   \begin{center}
      \caption[Sources of globular cluster photometry]{Sources of photometric data of globular clusters.}
      \label{clustasources}
         \begin{tabular}{llll}
            \hline
 cluster, {\rm [Fe/H]}, $(m-M)_{0}$, $E_{B-V}$ & colour bands &  source & range (in $m_{V}$) \\
            \hline
 47 Tuc (NGC 104), --0.70, 13.14, 0.04 & $B-V$       & \citet{wysocka}       & $\stackrel{_<}{_\sim} 19^{m}$ \\
                                      & $B-V$        & \citet{hesser}        & $\sim 24^{m}$ - $\sim 19^{m}$ \\
                                      & $V-I$        & \citet{barbuy}        & $\stackrel{_<}{_\sim} 13.5^{m}$ \\
                                      & $V-I$        & \citet{wysocka}       & $\sim 21^{m}$ -  $\sim 13.5^{m}$ \\
                                      & $U-B$        & \citet{gcpd}$^{a}$    & $\sim 16^{m}$ - $\sim 12^{m}$ \\
                                      & $V-R^{b}$    & \citet{cathey}        & $\sim 16^{m}$ - $\sim 12^{m}$ \\
                                      & $VJK$        & \citet{montegriffo}   & $\stackrel{_<}{_\sim} 17^{m}$ \\
                                      & $H-K$        & \citet{frogel}        & $\stackrel{_<}{_\sim} 14.5^{m}$ \\
 M 5 (NGC 5904), --1.11, 14.32, 0.03  & $B-V$        & \citet{sandquist}     & $\stackrel{_<}{_\sim} 22^{m}$ \\
                                      & $B-V$        & \citet{richer}        & $\sim 24^{m}$ - $\sim 22^{m}$ \\
                                      & $V-I$        & \citet{sandquist}     & $\stackrel{_<}{_\sim} 22^{m}$ \\
                                      & $U-B$        & \citet{drissen},      & $\stackrel{_<}{_\sim} 19.5^{m}$ \\
                                      &              & \citet{vonbraun}      & \\
                                      & $V-R$        & \citet{vonbraun}      & $\stackrel{_<}{_\sim} 17.5^{m}$ \\
                                      & $J-K$        & \citet{persson}       & $\stackrel{_<}{_\sim} 17.5^{m}$ \\
 M 3 (NGC 5272), --1.34, 15.01, 0.01  & $B-V$        & \citet{ferraro}       & $\stackrel{_<}{_\sim} 22^{m}$ \\
                                      & $V-I$        & \citet{ferraro}       & $\stackrel{_<}{_\sim} 16^{m}$ \\
                                      & $V-I$        & \citet{bolte}         & $\sim 23^{m}$ - $\sim 16^{m}$ \\
                                      & $V-I$        & \citet{marconi}       & $\sim 27^{m}$ - $\sim 23^{m}$ \\
                                      & $U-B$        & \citet{gcpd}$^{c}$    & $\stackrel{_<}{_\sim} 21.5^{m}$ \\
                                      & $V-R^{b}$    & Arribas and           & $\stackrel{_<}{_\sim} 14.5^{m}$ \\
                                      &              & Mart\'{\i}nez-Roger (1987) & \\
                                      & $V-K$        & \citet{frogel}        & $\stackrel{_<}{_\sim} 15^{m}$ \\
                                      & $J-K$        & \citet{kim}           & $\stackrel{_<}{_\sim} 15^{m}$ \\
 NGC 6397, --1.82, 11.75, 0.18        & $B-V$        & \citet{kaluzny}       & $\stackrel{_<}{_\sim} 18.5^{m}$ \\
                                      & $B-V$        & \citet{alvarado}      & $\sim 21.5^{m}$ - $\sim 18.5^{m}$ \\
                                      & $U-B$        & \citet{alvarado}      & $\sim 20^{m}$ - $\sim 16.5^{m}$ \\
                                      & $U-B$        & \citet{gcpd}$^{d}$    & $\stackrel{_<}{_\sim} 16.5^{m}$ \\
                                      & $V-I$        & \citet{alvarado}      & $\stackrel{_<}{_\sim} 12.5^{m}$ \\
                                      & $V-I$        & \citet{alvarado}      & $\sim 16.5^{m}$ - $\sim 12.5^{m}$ \\
                                      & $V-I$        & \citet{king}          & $\sim 27^{m}$ - $\sim 16.5^{m}$ \\
                                      & $R-I$        & \citet{alcaino}       & $\stackrel{_<}{_\sim} 16.25^{m}$ \\
 M 92 (NGC 6341), --2.16, 14.53, 0.02 & $B-V$        & \citet{sandage}       & $\stackrel{_<}{_\sim} 21.5^{m}$ \\
                                      & $B-V$        & \citet{stetson}       & $\sim 24^{m}$ - $\sim 21.5^{m}$ \\
                                      & $V-I$        & \citet{vonbraun}      & $\stackrel{_<}{_\sim} 14^{m}$ \\
                                      & $V-I$        & \citet{bolte}         & $\sim 20^{m}$ - $\sim 14^{m}$ \\
                                      & $V-I$        & \citet{piotto}        & $\sim 28^{m}$ -  $\sim 20^{m}$ \\
                                      & $U-B$        & \citet{gcpd}$^{e}$    & $\stackrel{_<}{_\sim} 20.5^{m}$ \\
                                      & $V-K$        & \citet{frogel}        & $\stackrel{_<}{_\sim} 15^{m}$ \\
                                      & $JHK^{f}$    & \citet{cohen}         & $\stackrel{_<}{_\sim} 15^{m}$ \\
            \hline
         \end{tabular}
   \end{center}
   {\tiny $^{a}$\citet{alcaino,evans,norris,lee,demarque,hartwick,cannon,menzies,eggen} \\
   $^{b}$transformed to Johnson-Cousins using \citet{bessellb} \\
   $^{c}$\citet{johnson,sandageb,sandage} \\
   $^{d}$\citet{woolley,newell,cannon,alcain,vandenbergh} \\
   $^{e}$\citet{walker,sandageb,sandage,eggen} \\
   $^{f}$transformed to Johnson using \citet{besselld}}
   \end{table*}
   In order to carry out the colour-calibration in a metallicity-dependent way, we
   collected empirical photometry from the literature of the best-studied
   Galactic globular clusters, spanning wide ranges in metallicity and
   evolutionary stages. The clusters used were 47 Tuc, M5, M3, NGC6397 and M92,
   with the respective ${\rm [Fe/H]}$-values of $-0.70$, $-1.11$, $-1.34$,
   $-1.82$, and $-2.16$ \citep{carretta}. To reach the best possible coverage
   of the multi-colour ($UBVRIJHKL$) - magnitude - diagram of each cluster,
   the photometry was combined from numerous sources
   summarised in table~\ref{clustasources}.
   Where necessary, the photometry was brought onto the Johnson-
   Cousins system using colour-colour relations by Bessell and
   collaborators \citep{bessella,bessellb,bessellc,besselld}.
   Where no fiducial lines were given, they were drawn by eye.
   In some colours (mainly $B-V$ and $V-I$), different sources were found
   for the same part of the CMD of the same cluster. The fiducials derived
   from or given by such sources usually didn't differ from each other by
   more than a few hundredths of a magnitude. \\
   In order to put these colour-magnitude diagrams onto the intrinsic
   system, the $E_{B-V}$ and $(m$-$M)_{V}$ values from the Harris online
   catalog of globular cluster parameters \citep{harris} were used.
   The adopted distance scale is based on the luminosity level of the
   horizontal branch (HB), where the absolute magnitude of the HB
   was derived from an empirical, metallicity-dependent relation
   ($M_{V}(HB) = 0.15 {\rm [Fe/H]} + 0.80$). Harris estimates that the absolute
   uncertainty of the predicted $M_{V}(HB)$ is of the order of 0.1-$0.2^{m}$,
   in the non-extreme cases of the used clusters more likely $0.1^{m}$.
   For the extinction coefficients, he gives an uncertainty of 
   $\Delta E_{B-V} = 0.1 E_{B-V}$. This shouldn't pose any problems, because we
   used clusters with extinction coefficients of only a few hundredths of a
   magnitude (only NGC 6397 has an $E_{B-V}$ of 0.18).
   The result are the fiducials shown in figs.~\ref{CMDsa} and \ref{CMDsb}
   (only the main sequence, the subgiant branch
   and the red giant branch are shown because the other parts of the
   CMD weren't used for the calibration). \\
   Two properties of the empirical CMDs are very striking. First, there are
   large gaps in the data. Measurements down to a few magnitudes below the main
   sequence turn-off exist only in $BVI$. In the far infrared, hardly any
   data at all are available. In order to create a sufficient photometric
   basis for population synthesis, more work clearly needs to be done on
   the observational side, in order to improve results based on either
   observational or synthetic spectra.\\
   The second striking point about the CMDs shown in figs.~\ref{CMDsa} and \ref{CMDsb} is the fact
   that their behaviour with increasing metallicity isn't as systematic or
   monotonous as one would expect.
   The colours of the RGB at a given ($V$) magnitude can differ up to
   $\sim 0.1^{m}$ from any trend with ${\rm [Fe/H]}$.
   The scatter around trends with metallicity is not of the same amplitude
   for all colours. While in $U-B$, $B-V$, and $R-I$ it is the most pronounced,
   $V-I$ behaves much more systematically with ${\rm [Fe/H]}$, as is also seen
   in homogeneous databases like the one from \citet{saviane}\footnote{In the infrared
   colours, there were barely enough data to derive a trend, but certainly not
   enough to determine the amplitude of a possible scatter around it.}.
   Part of the scatter (a few hundredths of a mag, see above) can be explained
   by the heterogeneity of the database which however cannot account for the
   entire scatter in those colours where it is observed.
   Obviously, there are parameters apart from age and metallicity that govern
   the appearance of the CMD. It is therefore more than amazing how in the
   literature, one often finds perfect agreement between synthetic isochrones
   and empirical CMDs for whole sets of globular clusters at the same time. \\
   From these CMDs, combined multi-colour ($UBVRIJHKL$) - ${\rm [Fe/H]}$ -
   ${T_{\rm eff}}$ - $\log g$ relations were synthesised, using the
   ${T_{\rm eff}}$ - ($V-K$) relation from the BaSeL 2.2 library. This relation
   incorporates \citet{ridgway} complemented with the
   differential properties in ${T_{\rm eff}}$ for solar metallicity of the
   original (uncalibrated) grid, as the latter is the only
   relation which covers the entire range in temperature.
   This resulting relation was used for all metallicities, because the $V-K$
   colour is expected to be metallicity insensitive
   \citep{vonbraun,martinez}.
   Finally, $\log g$ values were added to the calibration files, using
   empirical ${T_{\rm eff}}$ - $\log g$ relations for red giants from Cohen,
   Frogel and Persson \citep[][${\rm [Fe/H]}$-dependent]{cohen,frogel,persson}
   and for dwarfs from \citet[][${\rm [Fe/H]}$-independent]{angelov}.

   \begin{table*}
   \begin{center}
      \caption[Temperature ranges of empirical colour-${T_{\rm eff}}$ relations]{Temperature ranges of empirical colour-${T_{\rm eff}}$ relations.}
      \label{ranges}
         \begin{tabular}{lrll}
            \hline
Luminosity class & ${\rm [Fe/H]}$ & Colours & temperature range [K] \\
            \hline
giants & --0.5 & $UBVRIJHK^{a}$ & 3350 -  6000 \\
       & --1.0 & $UBVRIJHK$     & 3750 -  6000 \\
       & --1.5 & $UBVRIJHK$     & 3750 -  6000 \\
       & --2.0 & $UBVRIJHK$     & 4000 -  6000 \\
       &   all & $K-L$          & -            \\
dwarfs &   all & $UBVI^{b}$     & 4500 - 10000 \\
       &   all & $RJHKL^{c}$    & -            \\
            \hline
         \end{tabular}
   \end{center}
   {\tiny $^{a}UBVRIJHK$: $U-B$, $B-V$, $V-I$, $R-I$, $V-K$, $J-H$, $H-K$, $J-K$ \\
      $^{b}UBVI$: $U-B$, $B-V$, $V-I$ \\
      $^{c}RJHKL$: $R-I$, $V-K$, $J-H$, $H-K$, $J-K$, $K-L$}
         \end{table*}
   We could only produce colour-temperature relations in the ranges specified in
   Table~\ref{ranges}. For giants, the ranges include the entire RGB (except in
   $K-L$). Outside these ranges, we used the differential properties with
   regard to ${T_{\rm eff}}$ of the semi-empirical (BaSeL 2.2) grid, as there
   was no other information available (neither empirical nor theoretical) about
   the behaviour of these colours for these temperatures.
   In $K-L$ for all stars, and for dwarfs in $R-I$, $V-K$, $J-H$, $H-K$, and
   $J-K$, where no observed data were available, but which
   are needed to complete the calibration files, we had no choice but to adopt
   the ${T_{\rm eff}}$ - $K-L$ relation directly from the BaSeL 2.2 library. \\
   This set (i. e. metallicity-dependent colour - temperature - $\log g$)
   was complemented with the solar relations synthesised by Lejeune et al.
   to calibrate the BaSeL 2.2 library (see Paper II).
   \begin{figure*}
\includegraphics[width=\textwidth]{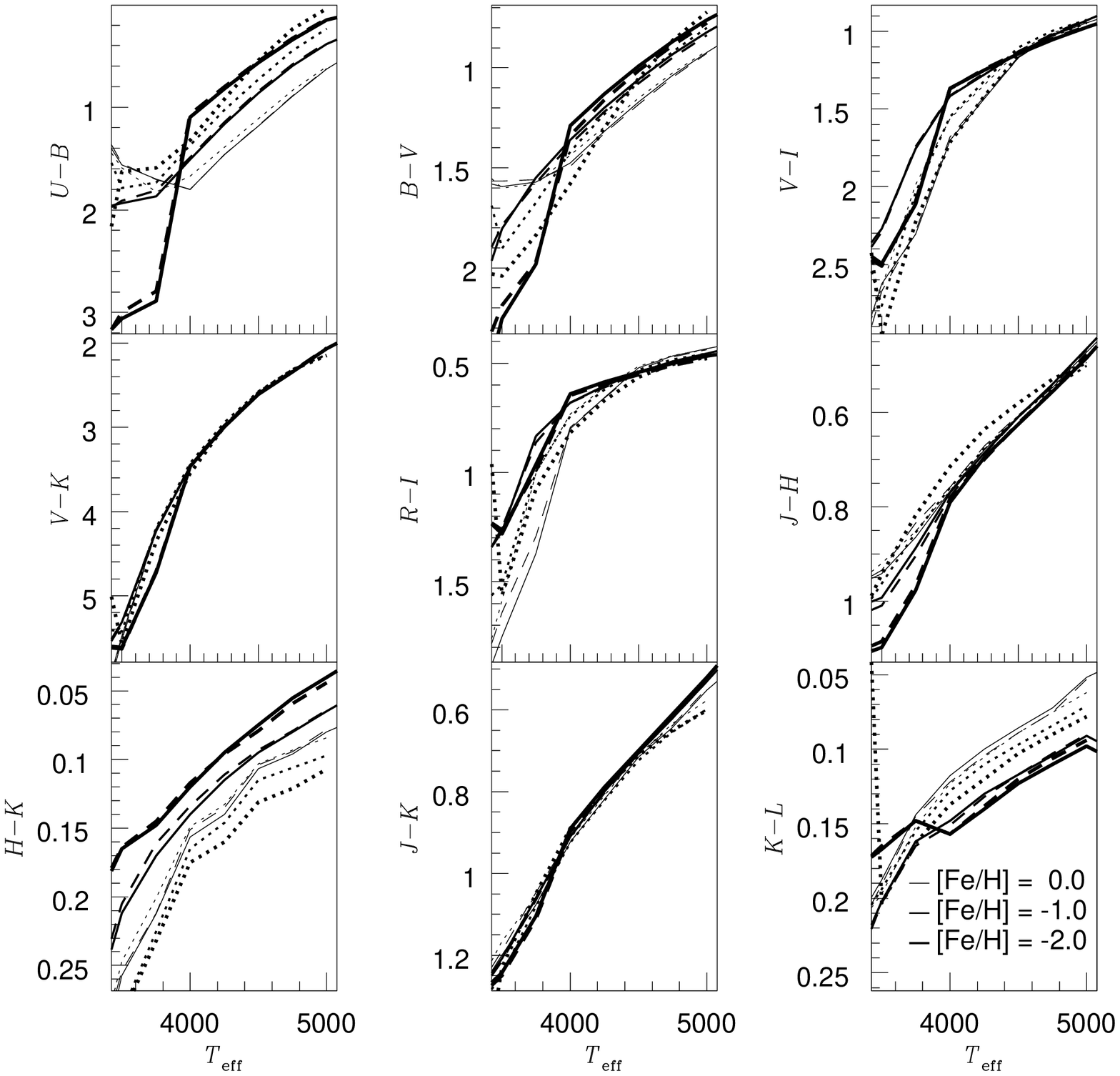}
   \caption{Empirical colour - temperature relations for giants
               of three grid metallicities (solid lines, increasing line width
               means decreasing metallicity) versus the relations from the
               BaSeL 3.1 ''WLBC 99'' models (dashed) resp. the
               BaSeL 2.2 semi-empirical models (dotted) for the same
               parameters.}
    \label{WLBC99coltemprelsgi}
    \end{figure*}
   \begin{figure*}
\includegraphics[width=\textwidth]{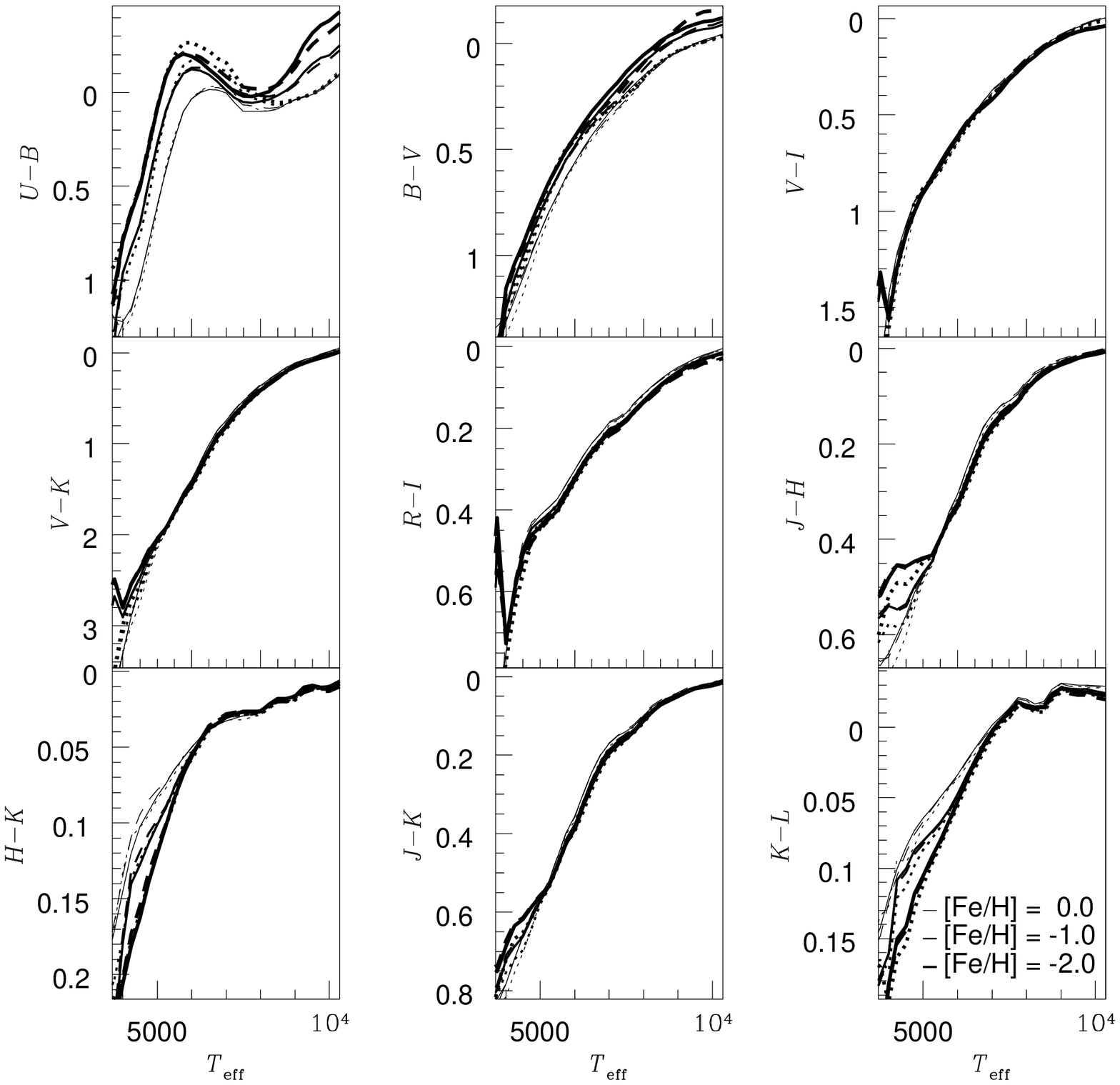}
   \caption{Empirical colour - temperature relations for dwarfs
               of three grid metallicities (solid lines, increasing line width
               means decreasing metallicity) versus the relations from the
               BaSeL 3.1 ''WLBC 99'' models (dashed) resp. the
               BaSeL 2.2 semi-empirical models (dotted) for the same
               parameters.}
    \label{WLBC99coltemprelsdw}
    \end{figure*}
   The so-derived calibration files (two for each metallicity
   (${\rm [Fe/H]} = 0, -0.5, -1, -1.5, -2$), one for (red) giants and one
   for dwarfs) are available by public ftp from the university of Basel.

\section{Metallicity-dependent library calibration, the BaSeL 3.1 ''WLBC 99'' library}
   These calibration files were then used to colour-calibrate the hybrid
   library of theoretical SEDs (consisting of the Kurucz-, the Allard
   and Hauschildt-, and the Scholz-models \citep{kurucz,allard,scholz}) using
   the (modified) algorithm developed by Cuisinier et al. (1997, see Paper I).
   For each temperature, it calculates a correction function which must
   be multiplied with the theoretical spectrum corresponding to the 
   parameters given in the calibration file for that temperature, such that the
   corrected spectra reproduce the colours given in the calibration file.
   This correction function being smooth, it leaves line strength indices
   untouched and only changes the continuum. This correction function is then
   used to correct all model spectra of this temperature, whereby the original
   differential properties with $\log g$ and ${\rm [Fe/H]}$ are usually
   conserved, and hence, continue to be well represented by the models.
   In the end, all SEDs are rescaled to make the bolometric fluxes match the
   temperatures (for a more precise description, see Paper I).
   Thus, the new feature of the calibration algorithm is to waive this
   differential approach in favour of calculating a separate correction
   function each temperature and for each level of ${\rm [Fe/H]}$, the
   differential properties with $\log g$ of the models are preserved, however.
   On top of that, the calibration programs were modified to include a smooth
   transition between the giant-calibrated and the dwarf-calibrated range. \\
   In this way, we created a metallicity-calibrated library of synthetic stellar
   SEDs, designed to reproduce observed colour-colour - and colour-temperature
   relations, which (for historical reasons) will from now on be called the
   BaSeL 3.1 ''WLBC 99'' library. \\
   As a first test of the new SED library, it was used to derive spectra and
   colours from the Yale and the Padova isochrones for the grid metallicities and
   for age 0 and 14 Gyr and also for the observed cluster metallicities
   at age 14 Gyr.
   As can be seen from figs.~\ref{WLBC99coltemprelsgi} and \ref{WLBC99coltemprelsdw} - where the input (calibration) relations
   (solid) are shown against the output (model) relations (dashed) and the
   BaSeL 2.2 semi-empirical relations (dotted) - the used colour - temperature relations
   are well-reproduced. This mainly confirms the functionality of the
   calibration algorithm, but also suggests that the colour - temperature
   relations of the models can be trusted.
   Because these relations have been calibrated in a metallicity-dependent way,
   the present results show improvements over earlier results obtained from the
   semi-empirical (BaSeL 2.2) library calibration in $U-B$, $J-H$, $H-K$
   and $J-K$, especially at low temperatures. They also appear smoother than
   their predecessors. Below the lower temperature limits of the
   calibration relations specified in table~\ref{ranges},
   they still show the same discontinuities as
   the BaSeL 2.2 models, due to lack of data, but for population synthesis,
   this shouldn't pose major problems, as stars of these low temperatures
   either don't show up at all (giants), or contribute only negligibly to the
   integrated light of a population (dwarfs).

   \begin{figure}
   \centering
   \includegraphics[width=\columnwidth]{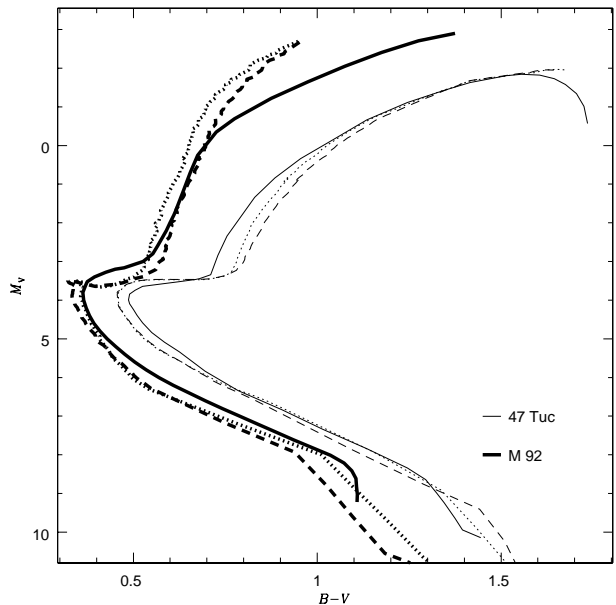}
   \caption{Empirical $(B-V)$ - $M_{V}$ CMD of the globular clusters 47 Tuc and M92
               (solid). Overlaid are the  CMDs created by combining the
               Padova 10 Gyr isochrone for ${\rm [Fe/H]} = -0.70$
               and the 16 Gyr isochrone for ${\rm [Fe/H]} = -2.16$
               with the BaSeL 3.1 ''WLBC 99'' library (dashed) resp. with
               the BaSeL 2.2 semi-empirical library (dotted).}
    \label{WLBC99isochronesBV}
    \end{figure}
   \begin{figure}
   \centering
   \includegraphics[width=\columnwidth]{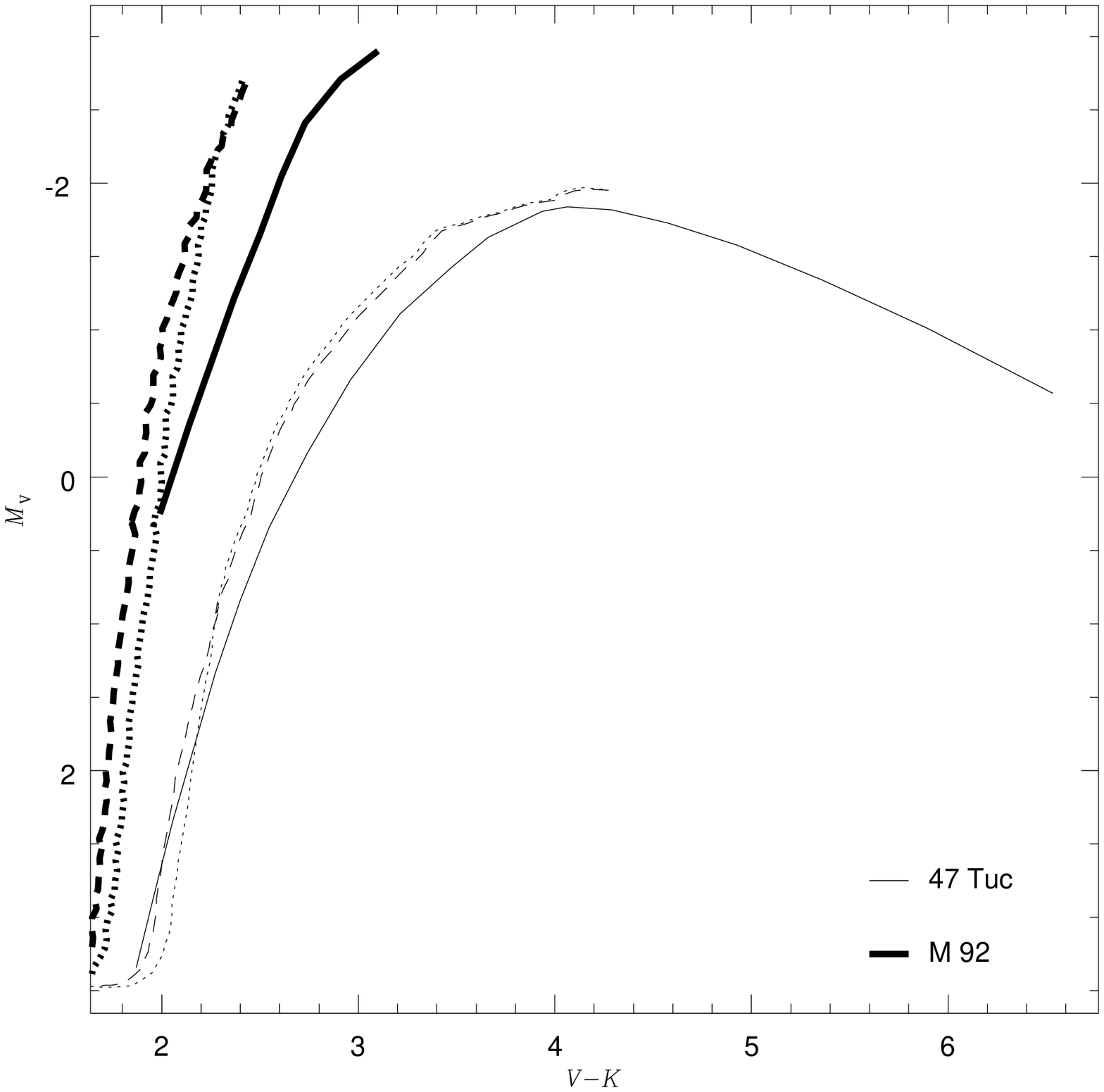}
   \caption{Empirical $(V-K)$ - $M_{V}$ CMD of the globular clusters 47 Tuc and M92
               (solid). Overlaid are the  CMDs created by combining the
               Padova 10 Gyr isochrone for ${\rm [Fe/H]} = -0.70$
               and the 16 Gyr isochrone for ${\rm [Fe/H]} = -2.16$
               with the BaSeL 3.1 ''WLBC 99'' library (dashed) resp. with
               the BaSeL 2.2 semi-empirical library (dotted).}
    \label{WLBC99isochronesVK}
    \end{figure}
   The second test unfortunately yields much less satisfying results. $M_{V} - 
   UBVRIJHK$ CMDs were produced using Yale and Padova isochrones for the
   observed metallicities of the calibrating globulars and typical cluster ages
   (eg. 10, 12, 12, 14 and 16 Gyr for 47 Tuc, M 5, M 3, NGC 6397, and M 92
   respectively), in order to reproduce
   their CMDs shown in figs.~\ref{CMDsa} and \ref{CMDsb}. Contrary to expectations, these isochrones
   do not reproduce the cluster CMDs better than the BaSeL 2.2 semi-empirical models.
   The RGBs come out too steep and in certain colours ($U-B$, $B-V$, $V-I$)
   too red for high metallicities, and too blue in all colours for low 
   values of ${\rm [Fe/H]}$ (see figs.~\ref{WLBC99isochronesBV} and
   \ref{WLBC99isochronesVK}). \\
   Obviously, it is impossible at the present epoch to establish a unique
   calibration of $UBVRIJHKL$ colours in terms of stellar metallicity that
   reproduces both empirical colour-temperature relations of stars and the CMDs
   of stellar populations (using theoretical isochrones).
   Whether these discrepancies are due to the used colour - temperature
   relations, or to the isochrones remains to be investigated, but recent
   indications point towards shortcomings in the isochrones, especially
   in the convection treatment \citep{salarisb}. \\
   Figs.~\ref{WLBC99coltemprelsgi}, \ref{WLBC99coltemprelsdw}, \ref{WLBC99isochronesBV}, and \ref{WLBC99isochronesVK} are only shown for Padova isochrones, but the results
   also hold for Yale isochrones, and the corresponding figures look very similar.
   The SED library is available by public ftp from the university of Basel as
   the BaSeL 3.1 ''WLBC 99'' version. Because of its good agreement for high
   metallicities, the (unmodified) ${\rm [Fe/H]}=+0.5$ file from the BaSeL 2.2
   library has been added to extend the metallicity range.

\section{Pragmatic solution: application-oriented library calibration,
the BaSeL 3.1 ''Padova 2000'' library}
   As obviously both the colour-temperature relations of stars and the CMDs
   of stellar populations (using existing theoretical isochrones) cannot be
   matched at the same time, we decided to also produce a library that is able
   to reproduce the CMDs of globulars well, because such a library should at
   least yield reliable integrated colours for stellar populations and hopefully
   reliable integrated spectra. Another motivation was the fact that according
   to some authors \citep{brocato}, colour-temperature relations, in particular
   for cool stars, aren't that well-known anyway. These authors even
   consider them a free parameter. \\
   As a theoretical isochrone library, which, combined with our to-be-produced
   SED library, should reproduce the CMDs of our calibration globular clusters,
   we chose the Padova 2000 isochrones \citep{padova}, because it is widely in
   use nowadays.

   We produced this library in an iterative process that is based on the fact
   (observed by us), that the $M_{V}$ magnitude, derived for a set of parameters
   ${\rm [Fe/H]}$, ${T_{\rm eff}}$, and $\log g$ from a stellar library,
   is practically independent of the choice of the spectral library. This is
   easily explained by the fact that $M_{bol}$ depends only on the stellar
   parameters (thus is independent of the choice of library), so the difference
   in $M_{V}$ stems only from the difference in the bolometric correction
   $BC(V)$ derived from the different libraries, which is negligible on the
   scale of absolute magnitudes.
   This library independence of $M_{V}$ can now be used to assign the colours
   of one's choice to a certain $M_{V}$ value, by assigning them to the
   ${T_{\rm eff}}$ and $\log g$ values that will reproduce this $M_{V}$ value
   in the to-be-calibrated library, which one knows already from an
   existing library. This way one can in principle shape a synthetical
   colour-magnitude diagram in the way one desires\footnote{Note that this
   procedure implies giving up consistency with empirically established
   ${T_{\rm eff}}$-colour relations.}. \\
   The iterative process employed was the following:
   \begin{enumerate}
      \item New calibration files were created from the ${T_{\rm eff}}$-$\log g$
         relations of the Padova 2000 isochrones for 0 Gyr for dwarfs (the age
         at which the main sequence is still complete) and 14 Gyr for giants
         (a typical globular cluster age), and colour-temperature
         relations derived via the colour-$M_{V}$ relations from the globular
         cluster fiducials presented in Section~\ref{compilation} and the
         $M_{V}$-${T_{\rm eff}}$ relations derived with the before-mentioned
         Padova 2000
         isochrones from the previous library (in the first step, the ''previous
         library'' was the ''WLBC 99'' library, and in the following steps,
         it was the library produced in the previous step).
         For solar metallicity, the calibration file was kept the same as
         for the ''WLBC 99'' library (thus also the same as for the BaSeL 2.2
         library), with the small difference, that the ${T_{\rm eff}}$-$\log g$
         relation was taken from the Padova 2000 isochrones for consistency.
       \item These calibration files were used to calibrate a new library.
       \item Steps 1 and 2 were repeated until there was no significant
          difference anymore between subsequent libraries.
   \end{enumerate}
   After four iteration steps this was the case. [The calibration files used
   for the final iteration are also available by ftp from Basel university,
   although they are probably of little use to the user].

   \begin{figure*}
\includegraphics[width=\textwidth]{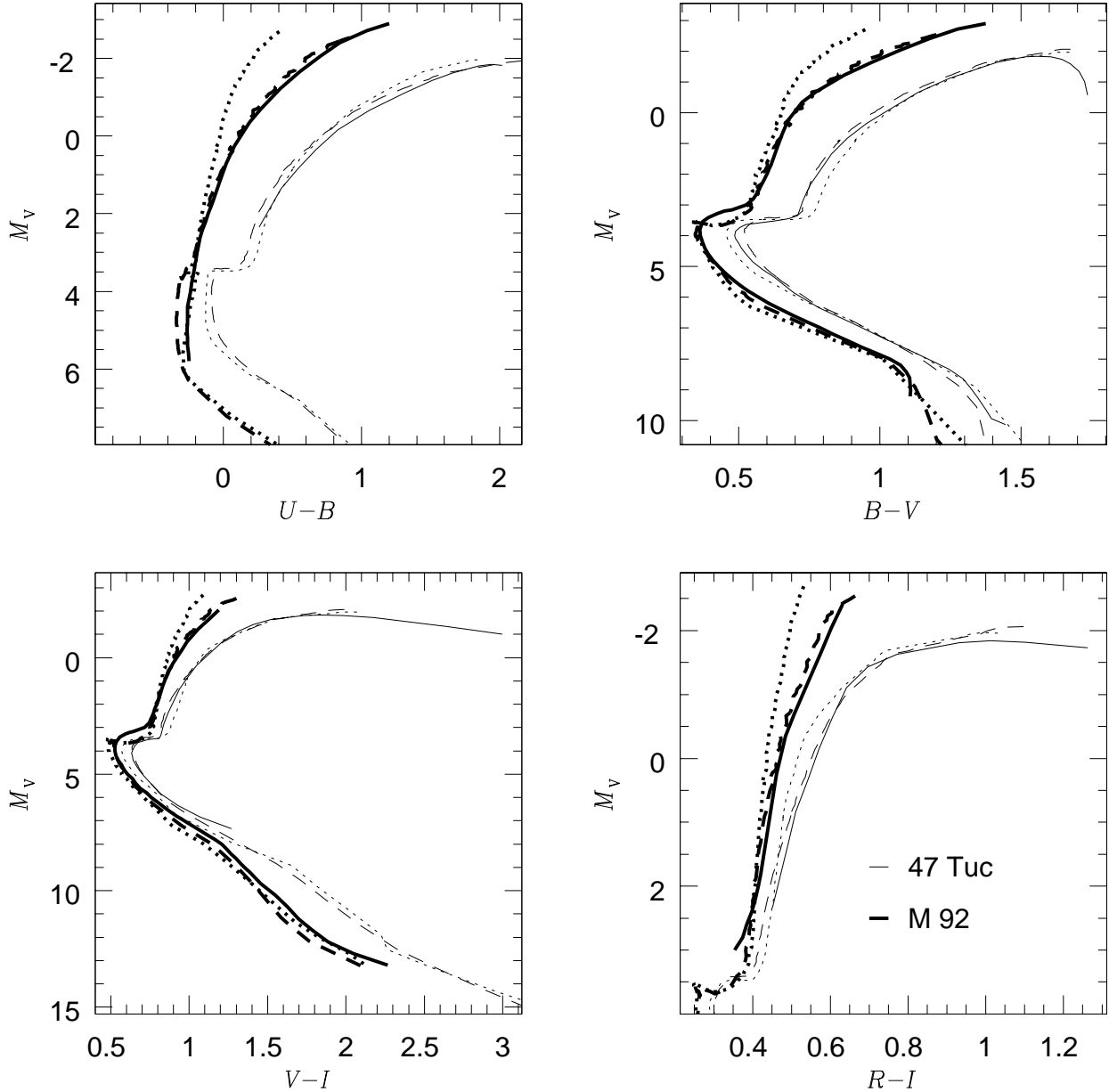}
   \caption{Empirical ($UBVRI$) CMDs of the globular clusters 47 Tuc and M92
               (solid). Overlaid are the  CMDs created by combining
               the Padova 10 Gyr isochrone for ${\rm [Fe/H]} = -0.70$
               and the 16 Gyr isochrone for ${\rm [Fe/H]} = -2.16$
               with the BaSeL 3.1 ''Padova 2000'' library (dashed) resp.
               with the BaSeL 2.2 semi-empirical library (dotted).}
    \label{PAD2000isochronesa}
    \end{figure*}
   \begin{figure*}
\includegraphics[width=\textwidth]{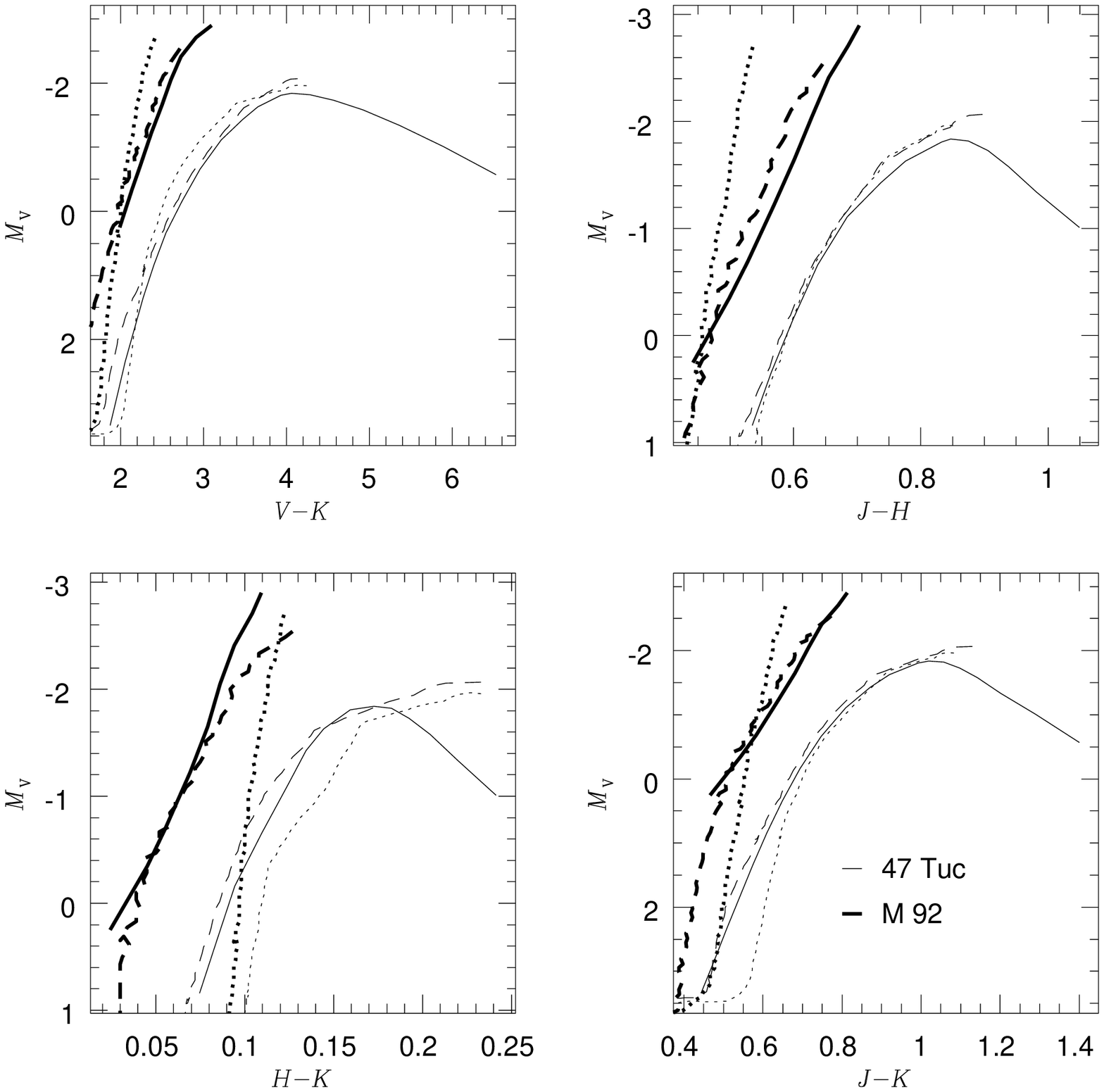}
   \caption{As in fig.~\ref{WLBC99coltemprelsgi}, but for infrared colours. The steepening in $J-H$, $H-K$
               and $J-K$ of the M 92 RGB above $M_{V} = \sim -1$ was deliberately
               not reproduced, as it's probably not a common RGB feature
               (maybe the tip of this fiducial was actually produced by
               asymptotic giant branch stars).}
    \label{PAD2000isochronesb}
    \end{figure*}
   As can be seen from figs.~\ref{PAD2000isochronesa} and \ref{PAD2000isochronesb}, the above procedure indeed provides the
   expected improvements.
   They show the globular cluster CMDs of figs.~\ref{CMDsa} and \ref{CMDsb} compared to the
   Padova 2000 isochrones of the same (cluster) metallicities and the
   above-mentioned ages evaluated with the new BaSeL 3.1 ''Padova 2000''
   library. \\
   The (metallicity-dependent) shapes and locations of the RGB and the main
   sequence are well-reproduced for the entire range from ${\rm [Fe/H]}$ =
   $-2.16$ to $-0.70$, apart from the RGB tip of 47 Tuc, of which the ''bend-down''
   in $V$ proved virtually impossible to reproduce
   (for solar ${\rm [Fe/H]}$, the quality of the library has already been
   confirmed extensively, as in its present form, it is almost identical
   with the \citet{paperii} BaSeL 2.2 library, which is widely used).
   At a given metallicity, the CMDs of individual clusters, however, can still
   differ by as much as 0.1$^{m}$ in colour, due to the above-mentioned large
   intrinsic cluster-by-cluster scatter in the observed CMDs.
   From the generally good overall agreement, it can be concluded that this new library
   should be useful for population synthesis, if combined with Padova 2000
   isochrones or tracks. It can be retrieved by ftp from Basel university as
   the BaSeL 3.1 ''Padova 2000'' library. Here too, the unmodified
   ${\rm [Fe/H]}=+0.5$ file from the BaSeL 2.2 library  was added.

   \begin{figure*}
\includegraphics[width=\textwidth]{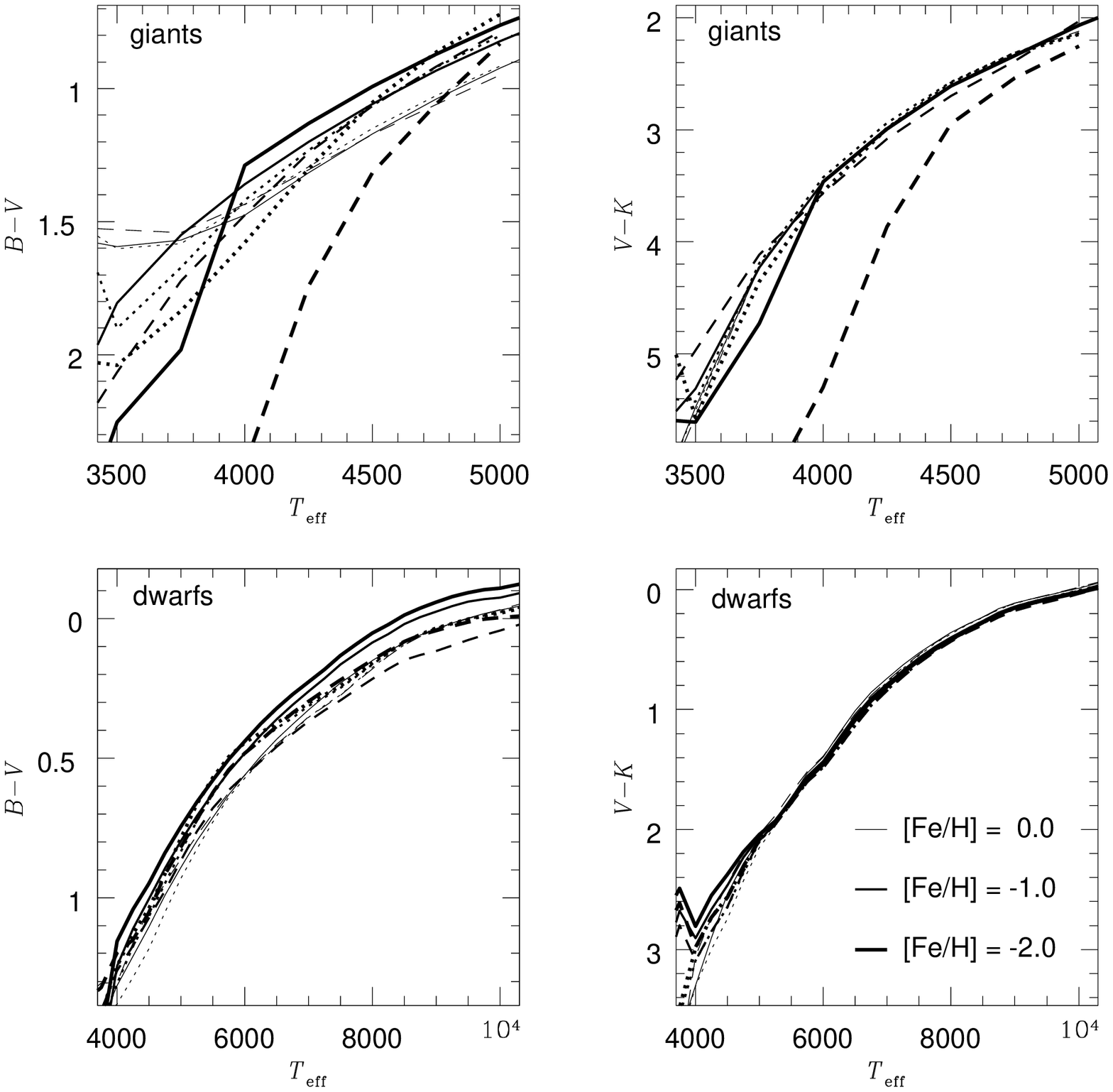}
   \caption{Empirical colour - temperature relations for giants and dwarfs
               of three grid metallicities (solid lines, increasing line width
               means decreasing metallicity) versus the relations from the
               BaSeL 3.1 ''Padova 2000'' models (dashed) resp. the BaSeL
               2.2 semi-empirical models (dotted) for the same parameters.}
    \label{PAD2000coltemprels}
    \end{figure*}
   The good agreement in the CMDs comes at the price of rather unusual
   colour temperature - relations. Fig.~\ref{PAD2000coltemprels} shows the empirical
   ${T_{\rm eff}}$ -  $V-K$  relation from \citet{ridgway} and
   empirical ${T_{\rm eff}}$ - $\log g$ relations for red giants from Cohen,
   Frogel and Persson \citep[][${\rm [Fe/H]}$-dependent]{cohen,frogel,persson}
   and dwarfs from \citet[][${\rm [Fe/H]}$-independent]{angelov}
   for three metallicities from the grid (solid lines) versus the BaSeL 3.1
   ''Padova 2000'' (dashed) and the BaSeL 2.2 semi-empirical (dotted) relations
   for the same parameters.
   For giants, the disagreement reaches up to 500 K or 1$^{m}$ in $V-K$
   (for colours with a smaller baseline, this error can be scaled down
   accordingly); for the
   lowest metallicities and temperatures (${\rm [Fe/H]} \leq -1.5$ and
   ${T_{\rm eff}} \leq 4000$ K) it can even reach 800 K, but fortunately,
   these stars don't show up in the Padova 2000 isochrones. For dwarfs,
   these problems aren't as bad as for giants, but it is clear that
   temperatures derived from the BaSeL 3.1 ''Padova 2000'' models should
   be treated with scepticism.
   On the positive side, most of the discontinuities have disappeared
   in the colour temperature - relations.

   \begin{figure*}
\includegraphics[width=\textwidth]{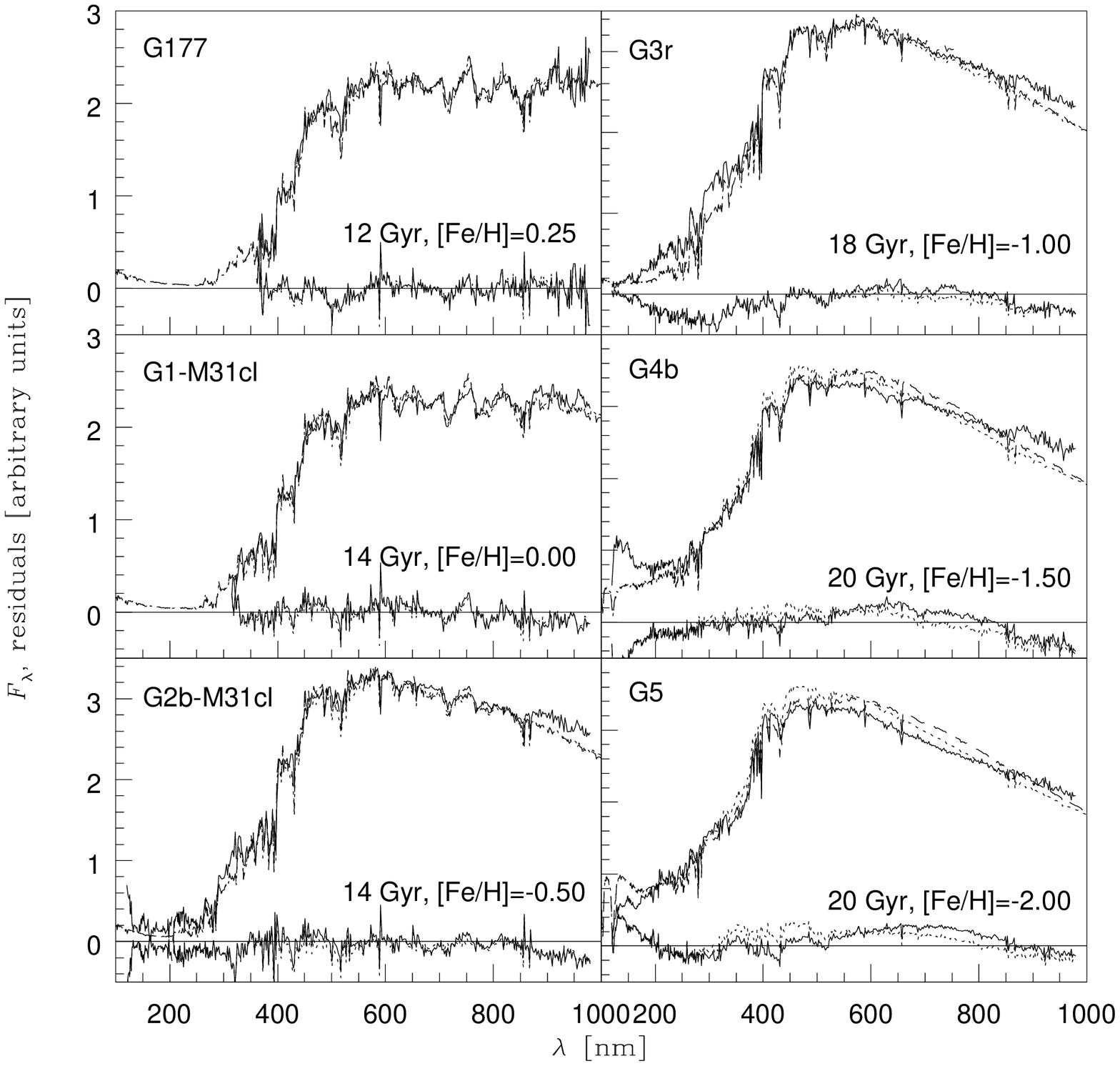}
   \caption{Template globular cluster integrated energy distributions
               from Bica et al. (1996, private communication) (solid lines)
               vs. integrated spectra of synthetic single burst stellar
               populations for the given ages and metallicities using
               the BaSeL 2.2 semi-empirical SED library (dotted) or
               the BaSeL 3.1 ''Padova 2000'' SED library (dashed).
               At the bottom of the figures, the residuals are shown,
               in the sense BaSeL 2.2 $-$ empirical spectra (dotted), or
               BaSeL 3.1 $-$ empirical spectra (solid). The zero-line is
               also shown in solid.}
    \label{spectra}
    \end{figure*}
%
   \begin{table}
   \begin{center}
      \caption[]{Colours for template globular cluster integrated energy distributions
               from Bica et al. (1996, private communication).}
      \label{bicacolours}
     $$ 
         \begin{array}{lrrrrr}
            \hline
            $template$ & ${\rm [Fe/H]}$ & U-B & B-V & V-I & R-I \\
            \hline
   $G177$      &  0.25 &      - &     - & 1.235 & 0.653 \\
   $G1-M31cl$  &  0.00 &      - & 0.993 & 1.238 & 0.663 \\
   $G2b-M31cl$ & -0.50 &  0.296 & 0.869 & 1.117 & 0.585 \\
   $G3r$       & -1.00 &  0.093 & 0.739 & 1.044 & 0.549 \\
   $G4b$       & -1.50 &  0.041 & 0.676 & 1.032 & 0.551 \\
   $G5$        & -2.00 & -0.001 & 0.639 & 0.948 & 0.507 \\
            \hline
         \end{array}
     $$ 
   \end{center}
   \end{table}
%
   \begin{table*}
   \begin{center}
      \caption[]{Magnitudes and colours for single stellar populations
                    sythesised using the BaSeL 3.1 ''Padova 2000'' SED library.}
      \label{padova2000colours}
     $$ 
         \begin{array}{lrrrrrrrrrrrrr}
            \hline
           $age$ & ${\rm [Fe/H]}$ & M_{bol} & M_{V} & U-B & B-V & V-I & V-K & R-I & J-H & H-K & J-K & K-L & K-M \\
            \hline
12.000 &  0.25 & 6.582 & 7.491 &  0.776 & 1.045 & 1.269 & 3.346 & 0.644 & 0.698 & 0.229 & 0.927 & 0.178 & 1.094 \\
14.000 &  0.00 & 6.616 & 7.407 &  0.599 & 0.978 & 1.224 & 3.130 & 0.634 & 0.691 & 0.200 & 0.892 & 0.164 & 1.020 \\
14.000 & -0.50 & 6.453 & 7.020 &  0.320 & 0.856 & 1.124 & 2.672 & 0.587 & 0.621 & 0.164 & 0.785 & 0.129 & 0.839 \\
18.000 & -1.00 & 6.523 & 6.976 &  0.155 & 0.771 & 1.051 & 2.366 & 0.532 & 0.555 & 0.135 & 0.691 & 0.120 & 0.735 \\
20.000 & -1.50 & 6.541 & 6.958 &  0.053 & 0.717 & 1.011 & 2.201 & 0.507 & 0.519 & 0.128 & 0.647 & 0.120 & 0.724 \\
20.000 & -2.00 & 6.563 & 6.978 & -0.009 & 0.684 & 0.980 & 2.135 & 0.498 & 0.499 & 0.121 & 0.620 & 0.116 & 0.731 \\
            \hline
         \end{array}
     $$ 
   \end{center}
   \end{table*}
%
   \begin{table}
   \begin{center}
      \caption[]{Differences in colours between
                    single stellar populations sythesised using 
                    the BaSeL 3.1 ''Padova 2000'' SED library and
                    template globular cluster integrated energy distributions
                    from Bica et al., in the sense BaSeL 3.1 $-$ template.}
      \label{residuals}
     $$ 
         \begin{array}{lrrrrr}
            \hline
            $template$ & ${\rm [Fe/H]}$ & \Delta U-B & \Delta B-V & \Delta V-I & \Delta R-I \\
            \hline
   $G177$      &  0.25 &      - &      - &  0.034 & -0.009 \\
   $G1-M31cl$  &  0.00 &      - & -0.015 & -0.014 & -0.029 \\
   $G2b-M31cl$ & -0.50 &  0.024 & -0.013 &  0.007 &  0.002 \\
   $G3r$       & -1.00 &  0.062 &  0.032 &  0.007 & -0.017 \\
   $G4b$       & -1.50 &  0.012 &  0.041 & -0.021 & -0.044 \\
   $G5$        & -2.00 & -0.008 &  0.045 &  0.032 & -0.009 \\
            \hline
         \end{array}
     $$ 
   \end{center}
   \end{table}
   In a last test, the effect of the metallicity-calibration on the integrated
   spectra of synthetic single-burst stellar populations was investigated.
   Synthetic populations were created in order to compare their integrated
   spectra with empirical template globular cluster integrated energy
   distributions from Bica et al. (1996, private communication).
   The synthetic populations were created using the GISSEL (Galaxy Isochrone Synthesis Spectral
   Evolution Library) software by G. Bruzual A. and S. Charlot
   \citep[2000, for a short description, see][]{charlot,bruzual} for 
   the template metallicities (${\rm [Fe/H]} = 0.25$, $0.00$, $-0.50$, $-1.00$,
   $-1.50$, $-2.00$), and the same ages as Lejeune used in his thesis (1997) for
   comparison of the BaSeL 2.2 semi-empirical library with the same template spectra
   (12, 14, 14, 18, 20 Gyr). Both the BaSeL 2.2 semi-empirical
   - and the BaSeL 3.1 ''Padova 2000'' SED libraries were used, along with
   a Salpeter IMF (mass range: $0.1 - 50 M_{\odot}$) and the Padova 2000
   tracks (except for ${\rm [Fe/H]} = -2.0$, where
   the Padova 1995 isochrone
   \citep{fagotto,girardi}
   was used, because the corresponding
   Padova 2000 isochrone wasn't implemented in the evolutionary
   synthesis program).
   The resulting integrated spectra are compared in fig.~\ref{spectra}, where the solid
   lines stand for the empirical template spectra, the dotted
   lines for the BaSeL 2.2 semi-empirical library and the dashed lines for the
   metallicity-calibrated library. Residuals in the sense BaSeL 2.2 $-$ empirical
   spectra (dotted), resp. BaSeL 3.1 $-$ empirical spectra (solid) are shown in
   the same figure. To guide the eye, the zero-line is also shown in solid.
   Synthetic ($UBVRIJHKLM$) colours and
   magnitudes from the empirical -  and the synthetical (BaSeL 2.2
   ''Padova 2000'') spectra are given in tables~\ref{bicacolours}, and
   \ref{padova2000colours}, and the differences between these colours in
   table~\ref{residuals} (due to the limited wavelength
   range of the Bica et al. spectra, only $UBVRI$ colours can be given for the
   empirical spectra).
   As expected, the differences between the two libraries are negligible for
   high ${\rm [Fe/H]}$. They are hardly visible in fig.~\ref{spectra},
   and in the colours, they are of the order of $0.001^{m}$.
   For low metallicities (${\rm [Fe/H]}=-1.65$), the metallicity-calibration
   seems to add flux in the red ($R$, $I$) and to reduce the flux
   in the visible a bit, but clearly, the dramatic differences
   in stellar colour-temperature relations between the two libraries
   have only a minor effect on the spectral properties of entire
   (synthetic) populations. The main improvements can be seen
   in the visible (the spectral shape in this region is matched perfectly
   for all metallicities), whereas in the ultraviolet one can still see
   some strong deviations.
   However, the differences between empirical integrated colours and the ones
   of the synthetic population only amount to a few hundredths of a magnitude
   (see table~\ref{residuals}), confirming that the BaSeL 3.1 ''Padova 2000'' SED library
   should be a useful tool for evolutionary synthesis, if used in combination
   with the Padova 2000 tracks/isochrones. \\

\section{Conclusions and future work}

   \begin{enumerate}
      \item We have constructed two comprehensive libraries of theoretical
         stellar energy distributions combining the Kurucz -, the Allard
         and Hauschildt -, and the Scholz-models
         \citep{kurucz,allard,scholz}. They provide synthetic stellar
         spectra with useful resolution on a homogeneous wavelength grid,
         from 9.1 nm to 160 $\mu$m, over large ranges of fundamental
         parameters. In addition to the preceding version, the BaSeL 2.2
         semi-empirical library (see Papers I and II), the present libraries
         have been ($UBVRIJHKL$) colour-calibrated independently at all levels
         of metallicity, using Galactic globular cluster photometric data, to
         overcome the weaknesses of the BaSeL 2.2 library at low ${\rm [Fe/H]}$.
      \item It proved impossible to calibrate the library in such a way that
         both the empirical colour-temperature relations (taken from
         \citet{ridgway}) and the CMDs of the calibrating clusters (using
         Yale or Padova 2000 isochrones) could be matched at the same time.
         Whether this is due to the isochrones or to irrealistic
         colour-temperature relations still remains to be investigated.
         Of great help would also be a more complete (colour) data basis.
         For this reason, we created two metallicity-calibrated libraries,
         the BaSeL 3.1 ''WLBC 99'' library,
         which is able to reproduce empirical colour-temperature relations
         and should therefore be a useful tool for investigations on a stellar
         level, and the BaSeL 3.1 ''Padova 2000'' library which, in combination
         with the Padova 2000 isochrones, successfully reproduces the CMDs
         (i. e. the location and shape of the different
         branches) of Galactic globular clusters, and which should therefore
         be a particularly suitable tool for spectral evolutionary synthesis
         studies of stellar systems and other synthetic photometry
         applications, especially if used in combination with the
         Padova 2000 isochrones/tracks.
      \item As the calibration algorithm is very flexible, new libraries
         can easily be produced as soon as better data and/or SED models
         are available. The present results, therefore, do not necessarily 
         provide the final version of the BaSeL models.
      \item As a first application, we are using the ''Padova 2000 tracks''
         library for the analysis of the colours of chemo-evolutionary
         dynamical models of spiral galaxies \citep{diss,westerasamland}.
   \end{enumerate}

\begin{acknowledgements}
      This work was supported by the Swiss National Science Foundation.
\end{acknowledgements}

\end{document}